\documentclass[12pt]{article}
\usepackage{amsfonts, amsmath, amssymb, cite, graphicx}

\usepackage{color}
\usepackage[all, knot]{xy}
\usepackage{tikz}
\usepackage{braket}

\usepackage[utf8]{inputenc}
\usepackage{appendix}
\usepackage{epstopdf}
\usepackage[footnotesize]{caption}
\usepackage{amsthm}
\usepackage{enumitem}
\usepackage{mathrsfs}

\usepackage[margin=3cm]{geometry}
\newcommand{\ba}{\begin{aligned}}
\newcommand{\ea}{\end{aligned}}

\def\be{\begin{equation}}
\def\ee{\end{equation}}
\def\bea{\begin{eqnarray}}
\def\eea{\end{eqnarray}}

\def\s{\sigma}

\def\l{\lambda}

\def\p{\partial}
\def\m{\mu}
\def\n{\nu}
\def\a{\alpha}

\def\>{\rangle} 
\def\<{\langle} 
\def\vev#1{\langle #1\rangle}

\def\tr{\text{tr}}

\def\vt1{\vartheta_1}
\def\vtv{\vartheta_\n}

\begin{document}
\begin{titlepage}
\vspace{0.5cm}
\begin{center}
{\Large \bf Correlation functions of the CFTs on torus with $T\bar{T}$ deformation}

\lineskip .75em
\vskip 2.5cm
{\large Song He$^{a,b,}$\footnote{hesong@jlu.edu.cn},  Yuan Sun$^{c,}$\footnote{sunyuan6@mail.sysu.edu.cn}}
\vskip 2.5em
 {\normalsize\it $^{a}$Center for Theoretical Physics and College of Physics, Jilin University, Changchun 130012, People's Republic of China\\
 $^{b}$Max Planck Institute for Gravitational Physics (Albert Einstein Institute),\\
Am M\"uhlenberg 1, 14476 Golm, Germany\\
$^{c}$   School of Physics and Astronomy, Sun Yat-Sen University, Guangzhou 510275, China}\\
\vskip 3.0em
\end{center}
\begin{abstract}
In this paper, we investigate the correlation functions of the conformal field theory (CFT) with the $T\bar{T}$ deformation on torus in terms of perturbative CFT approach, which is the extension of the previous investigations on correlation functions defined on a plane. We systematically obtain the first order correction to the correlation functions of the CFTs  with $T\bar{T}$ deformation in both operator formalism and path integral language. As a consistency check, we compute the deformed partition function, namely the zero-point correlation function, up to the first order, which is consistent with results in literature. Moreover,  we obtain a new recursion relation for correlation functions with multiple $T$'s and $\bar{T}$'s insertion in generic CFTs on torus. Base on the recursion relations, we study some correlation functions of stress tensors up to the first order under $T\bar{T}$ deformation.
  \end{abstract}
\end{titlepage}

\baselineskip=0.7cm

\tableofcontents
\newpage

\section{Introduction}
Recently a class of exactly solvable deformation of 2D QFTs with rotational and translational symmetries called $T\bar{T}$ deformation \cite{Zamolodchikov:2004ce,Smirnov:2016lqw,Cavaglia:2016oda} attracts a lot of research interest. With $T\bar{T}$ deformation, the deformed Lagrangian $\mathcal{L}(\l)$ can be written as
\be \label{TTLa}
\frac{\p \mathcal{L}(\l)}{\p \l}=-\int d^2z T\bar{T}(z),
\ee
where the composite operator $T\bar{T}(z)$ constructed from stress tensor within the theory $\mathcal{L}(\l)$ was first introduced in \cite{Zamolodchikov:2004ce}.
Although such kind of irrelevant deformation is usually hard to handle, it still has numerous intriguing properties. A remarkable property is integrability \cite{Smirnov:2016lqw,LeFloch:2019wlf,Jorjadze:2020ili}. If the un-deformed theory is integrable, there exists a set infinite of commuting conserved charges or KdV charges. After {$T\bar{T}$} deformation, these charges can be adjusted such that they still commute with each other \cite{Smirnov:2016lqw,LeFloch:2019wlf}. Hence in this sense the deformed theory is solvable. Furthermore, such deformation is well under control by the fact that it is possible to compute many quantities in the deformed theory especially when the un-deformed theory is a CFT, such as S-matrix, energy spectra, correlation functions, entanglement entropy and so on \cite{Cardy:2019qao,Rosenhaus:2019utc,Donnelly:2018bef,Chen:2018eqk,Sun:2019ijq,Jeong:2019ylz}. The $T\bar{T}$ deformation is a special one among a infinite set of deformations constructed from bilinear combinations of KdV currents \cite{Smirnov:2016lqw,LeFloch:2019wlf}. These deformations also preserve the integrability of the un-deformed theory. Besides $T\bar{T}$ deformation, other deformations in this set including the so-called $J\bar{T}$  deformation also receive  much attention from both field theory and holographic points of view \cite{Conti:2019dxg,Cardy:2018jho,Guica:2017lia,Apolo:2018qpq,Bzowski:2018pcy,Giveon:2019fgr,Chakraborty:2019mdf,Apolo:2019yfj,Jiang:2019trm}.  In addition, the $T\bar{T}$ deformation can also be understood from some other perspectives and generalizations \cite{Dubovsky:2017cnj,Cardy:2018sdv,Giveon:2017nie,Bonelli:2018kik,Baggio:2018rpv,Chang:2018dge,Jiang:2019hux,Chang:2019kiu,Coleman:2019dvf,Dubovsky:2018bmo,Conti:2018jho,Santilli:2018xux,Jiang:2019tcq,Giveon:2017myj,Asrat:2017tzd,Giribet:2017imm,Apolo:2019zai,Lewkowycz:2019xse}.

In particular, within $\l<0$, the  $T\bar{T}$-deformed CFT is suggested to be holographically dual to AdS space with Dirichlet boundary  condition imposed at finite radius \cite{McGough:2016lol,Guica:2019nzm}. On the boundary, the rotational and translational symmetries are still preserved, while the conformal symmetry is broken by the deformation. It opens a novel window to study holography without conformal symmetry. Many interesting progresses have been done along this direction, such as holographic entanglement entropy, holographic complexity etc. \cite{Shyam:2017znq,Kraus:2018xrn,Cottrell:2018skz,Bzowski:2018pcy,Taylor:2018xcy,Hartman:2018tkw,Shyam:2018sro,Caputa:2019pam,Apolo:2019yfj,Chen:2018eqk,Chen:2019mis,He:2019glx,Geng:2019yxo,Geng:2019ruz,Jafari:2019qns,Li:2020pwa}.

Correlation functions are fundamental observables in QFTs, so it is of great importance to study the correlation functions in its own right. The correlation functions have many important applications, e.g. quantum chaos, quantum entanglement, and so on. One example is the four-point functions which are related to out of time order correlation function (OTOC), a quantity that can be used to diagnose the chaotic behavior in field theory with/without the $T\bar{T}$ deformation \cite{Shenker:2014cwa, Roberts:2014ifa, Maldacena:2015waa,Apolo:2018oqv, He:2019vzf}. To measure the quantum entanglement, the computation of entanglement (or R\'{e}nyi) entropies involves the correlation functions \cite{Calabrese:2004eu}. In particular, the R\'{e}nyi entanglement entropy of the local excited states has been extensive calculated in various situations\cite{He:2014mwa,Guo:2015uwa,Chen:2015usa,He:2017vyf,He:2017txy,He:2017lrg,Guo:2018lqq}. In the present work we are interested in studying the correlation functions in the $T\bar{T}$ deformed CFT. In particular, the $T\bar{T}$ deformed partition function, namely zero-point correlation function, on torus could be computed and was shown to be modular invariant \cite{Datta:2018thy,Aharony:2018bad}. Furthermore the partition function
{with chemical potentials for KdV charges turning on} was also analyzed \cite{Asrat:2020jsh}. The correlation functions with $T\bar{T}$ deformation in the deep UV theory were investigated in a non-perturbtive way by J. Cardy \cite{Cardy:2019qao}.

Meanwhile, one can also proceed with conformal perturbation theory. Here we have to emphasize that we focus on the deformation region nearby the un-deformed CFTs, where the CFT Ward identity still holds and the effect of the renormalization group flow of the operator with the irrelevant deformation is not taken into account in the current setup. The conformal symmetry can be regarded as an approximate symmetry up to the lowest orders of the $T\bar{T}$ deformation and the correlation functions can be also obtained nearby the original theory. The total Lagrangian is expanded near the critical point for small coupling constant $\l$
\be
\mathcal{L}=\mathcal{L}_{CFT}-\l \int d^2z T\bar{T}(z).
\ee
The first order of deformed correlation functions take the following form
\be\label{eq1}
\l\int_{T^2} d^2z\vev{T\bar{T}(z)\phi_1(z_1)...\phi_n(z_n)},
\ee
where the expectation value in the integrand is calculated in the underformed CFTs by Ward identity, and the integration domain is the torus $T^2$.
In the perturbative CFTs approach, the deformed two-point functions and three-point functions were consider in \cite{Kraus:2018xrn,Guica:2019vnb} up to the first order in coupling constant. Subsequently, the present authors have considered the four-point functions \cite{He:2019vzf}. Also we generalized this study to the case with supersymmetric extension \cite{He:2019ahx}. Note that in the previous studies, these theories were defined on plane. In the present work, we would like to consider the theories defined on torus which will be very important to understand the boundary theory which is the holographic dual to the BTZ black hole \cite{KeskiVakkuri:1998nw}. The other motivation to study the correlation functions in the deformed theory on the torus is associated with reading the information about multiple entanglement entropy of the multi-interval \cite{Rajabpour:2015uqa,Rajabpour:2015nja,Rajabpour:2015xkj}. To obtain the deformed correlation functions, one has to calculate the integrand in eq.(\ref{eq1}) by Ward identity and do the integral over the torus $T^2$ with the help of a proper regularization scheme. The Ward identity on torus associated with the energy momentum tensor, e.g. $T$ or $\bar{T}$, has different structure compared with that on the plane \cite{He:2019vzf}\cite{He:2019ahx}. In terms of perturbative approach, we obtain the correlation functions with $T\bar{T}$ deformation systematically by using both operator formalism and path integral language following the analysis in \cite{Eguchi:1986sb,Felder:1989vx,w3}. In addition, the correlation functions in CFT with multiple $T$'s and $\bar{T}$'s insertion {can be also} obtained, for example, the case with a $T\bar{T}T\bar{T}$ insertion.

The plan of this paper is as follows. In section \ref{sec:TTbar}, we discuss the Ward identity associated with single $T$ and $\bar{T}$ insertion on torus and apply it to study the first order perturbation of partition function. Then we check the {partition functions} in the deformed free bosonic and fermionic field theories. {In section \ref{sec:nTTbar}, we obtained a recursion relations for multi $T$'s and $\bar{T}$'s inserted correlation functions in CFT, and apply it to the first order perturbation of the stress tensors correlation functions  under $T\bar{T}$ deformation. } In section \ref{sec:TTbarPI}, we offer the Ward identity on torus by using path integral method. Conclusions and discussions are given in the final section. In appendices, we would like to list the notations and some relevant techniques which are very useful in our analysis.

\section{$T\bar{T}$-deformation}\label{sec:TTbar}
In this section we will calculate the first order $T\bar{T}$ correction to the correlation functions eq.(\ref{eq1}) of the CFTs on torus. As examples, the results {are} applied to the first order corrections to the partition function in free field theories with $T\bar{T}$ deformation.
\subsection{Correlation functions in the $T\bar{T}$-deformed CFTs }
To obtain the correlation functions of the CFTs with $T\bar{T}$ {deformation} on torus, the procedure is the similar as the case in which there is only a single $T$-insertion\cite{Felder:1989vx,w3}, where the correlation functions were derived in the operator formalism. Interestingly, the same results were also obtained in path integral language \cite{Eguchi:1986sb}. We start with recalling the well-known result about the $T$ inserted correlation functions on torus  in CFTs\cite{DiFrancesco:1997nk}
\be\ba \label{Toperator}
&\vev{T(w)X}-\vev{T}\vev{X}\\
=&\sum_i \Big(h_i(P(w-w_i)+2\eta_1)+ (\zeta(w-w_i)+2\eta_1w_i)\p_{w_i}\Big)\vev{X}+ 2\pi i \p_\tau\vev{X},
\ea\ee
where $X\equiv \phi_1(w_1,\bar{w}_1)...\phi_n(w_n,\bar{w}_n)$, a string of primary operators,  $P(z),\zeta(z)$ are the Weierstrass $P$-function and zeta function respectively, $\eta_1=\zeta(1/2)$, and $\tau$ is modular parameter of the torus\footnote{For our conventions, please refer to appendix \ref{conventions}.}.  Although the prefactor $(\zeta(w-w_i)+2\eta_1 w_i)$ is not doubly periodic on coordinate $w$, the correlation function $\vev{T(w)X}$ is doubly periodic on $w$ by translation symmetry. In fact,  eq.(\ref{Toperator}) can be regarded as a generalization of Ward identity on plane. As $w\to w_i$, the usual OPE on the plane is reproduced
\be\ba
T(w)\phi_i(w_i,\bar{w}_i)\sim \frac{h_i\phi_i(w_i,\bar{w}_i)}{(w-w_i)^2}+\frac{\p_{w_i}\phi(w_i,\bar{w}_i)}{w-w_i},
\ea\ee
where we used the expansion of functions $P(w)\sim 1/w^2,\zeta(w)\sim 1/w$ in the neighborhood of point $w=0$.

In what follows, we will review how to derived eq.(\ref{Toperator}) in operator formalism as in \cite{Felder:1989vx}. The partition function on torus is defined by the following trace over the Hilbert space
\be
Z=\tr(q^{L_0-c/24}\bar{q}^{\bar{L}_0-c/24}),~~q=e^{2\pi i\tau}.
\ee
The correlation functions {of} $X(\{w_i,\bar{w}_i\})=\phi_1(w_1,\bar{w}_1)...\phi_n(w_n,\bar{w}_n)$\footnote{We will suppress the anti-holomorphic coordinates $\bar{w_i}$ {dependence} in $X$ for simplicity hereafter.} takes the form
\be
\vev{X(\{w_i\})}=\frac{1}{Z}\tr(X(\{w_i\})q^{L_0-c/24}\bar{q}^{\bar{L}_0-c/24}).
\ee
To obtain the $T$ inserted correlation function $\vev{T(w)X(\{w_i\})}$, we started with the coordinate $z$ on plane  which is related to standard coordinate $w$ on cylinder via the exponential map\footnote{We will refer to  coordinate $w$ as standard coordinate on torus.} $z=e^{2\pi i w}$. On the plane one can expand the stress tensor as
\be\ba \label{ExpT}
T_{pl}(z)=\sum_{n\in \mathbb{Z}}\frac{L_n}{z^{n+2}}.
\ea\ee
Now consider the quantity $
\tr(T_{pl}(z)X(\{z_i\})q^{L_0-c/24})$,\footnote{For simplicity we suppressed the anti-holomorphic factor $\bar{q}^{\bar{L}_0-c/24}$ inside the trace.}  using (\ref{ExpT}), which equals
\be\ba  \label{tpq}
&\tr(T_{pl}(z)X_{pl}(\{z_i\})q^{L_0-c/24})=\frac{1}{z^2}\tr(L_0X_{pl}q^{L_0-c/24})+\sum_{n\neq 0}\frac{1}{z^{n+2}}\tr(L_nX_{pl}q^{L_0-c/24}),
\ea\ee
where $X_{pl}(\{z_i\})$ are primary operators defined on plane. The first term can be converted to the derivative with respect to the modular parameter $\tau$
\be\ba
\tr(L_0X_{pl}q^{L_0-c/24})=\frac{1}{2\pi i}\frac{\p}{\p \tau}\tr(X_{pl}q^{L_0-c/24})+\frac{c}{24  }\tr(X_{pl}q^{L_0-c/24}),
\ea\ee
while the second term equals
 \footnote{An useful relation
\be\label{formula}
q^{L_0}L_nq^{-L_0}=q^{-n}L_n.
\ee}
\be \ba
\tr(L_nX_{pl}q^{L_0-c/24})=q^{-n}\tr(L_nq^{L_0-c/24}X_{pl})=\frac{1}{q^n-1}\tr(q^{L_0-c/24}[X_{pl},L_n]).
\ea\ee
Note the commutator on the RHS can be further expressed as a contour integral
\be
[L_n,X_{pl}(\{z_i\})]=\frac{1}{2\pi i}\oint_\gamma dz_0 z_0^{n+1}T(z_0)X_{pl}(\{z_i\}).
\ee
Here the contour $\gamma$ encircles the  operators located at $z_i,i=1,...,n$. Then eq.(\ref{tpq}) is
\be\ba\label{tpq2}
&\tr(T_{pl}(z)X_{pl}(\{z_i\})q^{L_0-c/24})\\
=&\frac{1}{z^2}\frac{1}{2\pi i}\frac{\p}{\p \tau}\tr(X_{pl}q^{L_0-c/24})+\frac{c}{24z^2}\tr(X_{pl}q^{L_0-c/24})\\&+\frac{1}{2\pi i}\oint_\gamma dz_0\frac{z_0}{z^{2}}\Big(-\frac{1}{2\pi i}\zeta(w_0-w)+\frac{1}{\pi i}\eta_1(w_0-w)-\frac{1}{2}\Big)  \tr(T_{pl}(z_0)X_{pl}q^{L_0-c/24}),
\ea\ee
where the following formula \cite{Felder:1989vx} is used
\be\label{for1}
\sum_{n\neq 0}\frac{1}{1-q^n}\Big(\frac{z_0}{z}\Big)^n=-\frac{1}{2\pi i}\zeta(w_0-w)+\frac{1}{\pi i}\eta_1(w_0-w)-\frac{1}{2}
\ee
with $z_0=e^{i2\pi w_0},z=e^{2\pi iw}$. Note the contour $\gamma$ does not encircle $z$.

Next we transform all the quantities above  on plane to coordinate $w$ on torus by exponential map.
For stress tensor on torus $T(w)$, one has
\be \label{TransT}
z^2 T_{pl}(z)=\frac{1}{(2\pi i)^2}T(w)+\frac{c}{24},
\ee
and the primary fields $X_{pl}(\{z_i\})$ transform accordingly to $X(\{w_i\})$ on torus.
It follows that eq.(\ref{tpq2}) can be written as
\be\ba  \label{wipre}
&\tr( T(w) X(\{w_i\})q^{L_0-c/12})\\=&2\pi i\frac{\p}{\p \tau}\tr( X(\{w_i\})q^{L_0-c/24})\\&+   \frac{1}{2\pi i}\oint_{\gamma'} dw_0  \Big(- \zeta(w_0-w)+2\eta_1(w_0-w)-\pi i\Big) \tr( T(w_0) X(\{w_i\})q^{L_0-c/24}),
\ea\ee
where the contour  on torus $\gamma'$ transformed from $\gamma$ on plane encloses $w_i$ not $w$.
 It can be shown that the above equation is also valid when $X$ contains component of the stress tensor $T$.
 The second term on the RHS can be further evaluated by substituting into the OPE
\be \label{OPETp}
T(w_0)\phi_i(w_i)\sim \frac{h_i\phi_i(w_i)}{(w_0-w_i)^2}+\frac{\p_i\phi_i(w_i)}{w_0-w_i},
\ee
which leads to
\be\ba \label{int4}
&\frac{1}{2\pi i}\oint_{\gamma'} dw_0  \Big(- \zeta(w_0-w)+2\eta_1(w_0-w)-\pi i\Big)\tr( T(w_0) X(\{w_i\})q^{L_0-c/24})\\
=
& \sum_i h_i\tr(q^{L_0-c/24}X)   \Big(- \zeta'(w_i-w)+2\eta_1 \Big) +   \Big(- \zeta(w_i-w)+2\eta_1 w_i\Big) \p_{w_i}\tr( X  q^{L_0-c/24}).
\ea\ee
where in the last step the translation symmetry is used ($\sum_i\p_{w_i}\vev{X}=0$). Finally we  obtain
\be\ba  \label{wiT}
&\tr( T(w) Xq^{L_0-c/12})-2\pi i\frac{\p}{\p \tau}\tr(Xq^{L_0-c/24})\\=&
\sum_i h_i \Big(- \zeta'(w_i-w)+2\eta_1 \Big) \tr(q^{L_0-c/24}X)  + \sum_i  \Big(- \zeta(w_i-w)+2\eta_1 w_i\Big) \p_{w_i}\tr( X q^{L_0-c/24}).
\ea\ee
After dividing  both side of eq.(\ref{wiT}) by $Z$, the result eq.(\ref{Toperator}) is produced.

Based on the derivation above, we can next consider $T\bar{T}$ insertion, which can be done by replacing $X$ in eq.(\ref{wipre}) with  $\bar{T}(\bar{v})X$. Since OPE $T$ with $\bar{T}$ is regular, only the OPE $T\phi_i$ will contribute to the contour integral. Following the same line as eq.(\ref{OPETp})-eq.(\ref{wiT}), the $T\bar{T}$ inserted correlation function is given by
\be\ba \label{ttb5}
&\tr( T(w)\bar{T}(\bar{v}) Xq^{L_0-c/12})\\=&2\pi i\frac{\p}{\p \tau}\tr(\bar{T}(\bar{v})Xq^{L_0-c/24}) +\sum_i h_i \Big(- \zeta'(w_i-w)+2\eta_1 \Big)\tr(q^{L_0-c/24}\bar{T}(\bar{v})X)   \\&+ \sum_i  \Big(- \zeta(w_i-w)+2\eta_1  w_i-2\eta_1 w-\pi i \Big) \p_i\tr(\bar{T}(\bar{v}) X q^{L_0-c/24}),
\ea\ee
where we have implicitly included the factor $\bar{q}^{\bar{L}_0-c/24}$ inside the trace. Equivalently,
\be\ba \label{ttb61}
&\vev{ T(w)\bar{T}(\bar{v}) X}\\
=&2\pi i \p_\tau\vev{\bar{T}(\bar{v})X }+2\pi i( \p_\tau\ln Z)\vev{\bar{T}(\bar{v}) X} +\sum_i h_i \Big(- \zeta'(w_i-w)+2\eta_1 \Big)\vev{\bar{T}(\bar{v})X}  \\& + \sum_i  \Big(- \zeta(w_i-w)+2\eta_1 w_i-2\eta_1 w-\pi i  \Big) \p_{w_i}\vev{\bar{T}(\bar{v})X}.
\ea\ee
Consider the term in last line $(-2\eta_1 w-\pi i)\sum_i\p_{w_i}\vev{\bar{T}(\bar{v})X}$, using translation symmetry, one has
\be \label{cont}
\sum_i\p_{w_i}\vev{\bar{T}(\bar{v})X}=-\p_{v}\vev{\bar{T}(\bar{v})X}.
\ee
Substituting the anti-holomorphic counterpart of eq.(\ref{Toperator}) into the RHS,  one can see that $ \p_{v}\vev{\bar{T}(\bar{v})X}$ is analytic on torus except at the contact points $v\sim w_i$. Explicitly, using
\footnote{The convention for delta function here is
$\p_{\bar{z}}\frac{1}{z}=\pi\delta^{(2)}(z),\delta^{(2)}(z)=\delta(x)\delta(y)$ where $z=x +i y$.
}
\be\ba  \label{derPZ}
&\p_v\bar{P}(\bar{v}-\bar{w}_i)\sim \p_v\frac{1}{(\bar{v}-\bar{w}_i)^2}=-\p_v\p_{\bar{v}} \frac{1}{\bar{v}-\bar{w}_i}=-\pi \p_{\bar{v}}\delta^{(2)}(v-w_i),\\
&\p_v\bar{\zeta}(\bar{v}-\bar{w}_i)\sim \p_v \frac{1}{\bar{v}-\bar{w}_i}=\pi  \delta^{(2)}(v-w_i),
\ea\ee
one can get
\be\ba
\p_{v}\vev{\bar{T}(\bar{v})X}=\pi\sum_i\Big( -h_i\p_{\bar{v}}\delta^{(2)}(v-w_i)+ \delta^{(2)}(v-w_i)\p_{\bar{w}_i}\Big) \vev{X},
\ea\ee
which means the last two terms in the last line of eq.(\ref{ttb61}) are contact terms vanishing on torus except at contact points. Following the prescription in \cite{Dijkgraaf:1996iy},  when computing the integral in the first order perturbation of $T\bar{T}$ deformed correlation functions, we excise these singular points $v=w_i$ from the integral domain
\be \label{firord}
\l\int_{T^2 -\sum_i D(w_i)} d^2v\vev{T(v)\bar{T}(\bar{v})X},
\ee
where $D(w_i)$ is a small disk centered at $v=w_i$. Therefore in this prescription the term $(-2\eta_1 w-\pi i)\sum_i\p_{w_i}\vev{\bar{T}(\bar{v})X}$ in the last line of eq.(\ref{ttb61}) make no contribution to the first order  $T\bar{T}$ deformed correlation functions.

It is interesting to apply the eq.(\ref{ttb61}) to the {case} where $X$ is identity operator
\be\ba  \label{vevttb}
&\vev{ T(w)\bar{T}(\bar{v}) }=2\pi i \p_\tau\vev{\bar{T}(\bar{v}) }+2\pi i \p_\tau\ln Z\vev{\bar{T}(\bar{v}) }=-(2\pi i)^2\frac{1}{Z} \p_\tau\p_{\bar{\tau}}Z,
\ea\ee
where we have used $\vev{\bar{T}(\bar{v}) }=-2\pi i\p_{\bar{\tau}}\ln Z$. The above result indicates {the} expectation value of $\vev{T\bar{T}}$ operator on torus does not dependent on the position $w,v$, which is reasonable due to the translation invariance. This can be seen more explicitly from {eq.(\ref{TTb0})} below that only zero modes of stress tensor  contribute to $\vev{ T(w)\bar{T}(\bar{v}) }$ and coordinate dependent terms vanish. The same phenomenon also presents in the cylinder case \cite{Zamolodchikov:2004ce}.

Actually  eq.(\ref{vevttb}) can be derived in a more direct way. To see this we start with the trace of a single insertion of stress tensor on plane
\be
\tr({T_{pl}(z)q^{L_0-\frac{c}{24}}})= z^{-2}\sum_n z^{-n}\tr(q^{L_0-\frac{c}{24}}L_n)=  z^{-2}\tr(q^{L_0-\frac{c}{24}}L_0),
\ee
where we used eq.(\ref{formula}) such that the terms with $n\neq 0$ vanish. Next transform that to torus by the map (\ref{TransT})
\be
\tr([\frac{1}{(2\pi i)^2}T(w)+\frac{c}{24}]q^{L_0-\frac{c}{24}} )=\tr(q^{L_0-\frac{c}{24}}L_0)=\frac{1}{2\pi i}\frac{\p }{\p \tau}\tr(q^{L_0-\frac{c}{24}})+\tr(\frac{c}{24}q^{L_0-\frac{c}{24}}).
\ee
The expectation value of $T$ is then obtained
\be \label{tnox}
\tr( T(w)  q^{L_0-\frac{c}{24}} ) = 2\pi i \frac{\p }{\p \tau}\tr(q^{L_0-\frac{c}{24}})
,~~\text{or}~~
\vev{T(w)}=2\pi i \frac{\p }{\p \tau}\ln Z.
\ee
Now consider $T(z_1)\bar{T}(\bar{z}_2)$ insertion, which is
\be\ba \label{ttbnox}
\tr(q^{L_0}\bar{q}^{\bar{L}_0}T(z_1)\bar{T}(\bar{z}_2))=&z_1^{-2}\bar{z}_2^{-2}\sum_{n,m}\tr(q^{L_0}\bar{q}^{\bar{L}_0}L_n\bar{L}_m)z_1^{-n}\bar{z}_2^{-m}.
\ea\ee
Noting $[L_n,\bar{L}_n]=0$ and using eq.(\ref{formula}), one has
\be\ba
\tr(q^{L_0}\bar{q}^{\bar{L}_0}L_n\bar{L}_m)=q^{-n}\tr(\bar{q}^{\bar{L}_0}L_nq^{L_0}\bar{L}_m)=q^{-n}\tr(q^{L_0}\bar{q}^{\bar{L}_0}L_n\bar{L}_m),
\ea\ee
thus
\be\ba\label{zeromode}
\tr(q^{L_0}\bar{q}^{\bar{L}_0}L_n\bar{L}_m)=\delta_{m0}\delta_{n0}\tr(q^{L_0}\bar{q}^{\bar{L}_0}L_0\bar{L}_0),
\ea\ee
which indicates only the term with $n=m=0$ will contribute to the summation in eq.(\ref{ttbnox}).
Further making transformation to torus and using eq.(\ref{tnox}), we finally obtain
\be\ba \label{TTb0}
\vev{T(w_1)\bar{T}(\bar{w}_2)}=-(2\pi i)^2\frac{1}{Z}\p_\tau\p_{\bar{\tau}}Z,
\ea\ee
which is the same as eq.(\ref{vevttb}).

It is interesting to note that the expectation value $\vev{T\bar{T}}$ is related to the  first order perturbation of partition function under $T\bar{T}$ deformation. The deformed partition function is
\be\ba
Z'=&\int D\phi e^{-S+\l\int d^2z T\bar{T}(z)}=Z(1+\l \int d^2z\vev{T\bar{T}}(z))...).
\ea\ee
with the CFT partition function $Z=\int D\phi e^{-S}$. By substituting  eq.(\ref{vevttb}), the first order perturbation of partition function is
\be
\l Z \int d^2z\vev{T\bar{T}}(z)=\l(2\pi )^2\tau_2\p_\tau\p_{\bar{\tau}}Z.
\ee
which is in good agreement with the result in \cite{Datta:2018thy}, where the partition function with $T\bar{T}$ deformation was computed by using the deformed spectrum. 
\subsection{Free field theories}
Now we apply the formula eq.(\ref{vevttb}) to free field theories, and show that eq.(\ref{vevttb})  is consistent with the results obtained by Wick contraction.

Let us first consider the free boson on torus. The   CFT partition function is
\be
Z(\tau)=\frac{1}{\sqrt{\tau_2}|\eta(\tau)|^2},
\ee
where $\eta(\tau)$ is the Dedekind $\eta$-function. The two-point function of scalar fields is well-known, which takes the form \cite{DiFrancesco:1997nk}\footnote{Note the coordinate $z_i$  in $\phi(z_i,\bar{z}_i)$ is standard coordinate on torus. }
\be
\vev{\phi(z_1,\bar{z}_1)\phi(z_2,\bar{z}_2)}=- \log\Big|\frac{\vartheta_1(z_{12}/2w_1)}{\eta(\tau)}\Big|^2+2\pi\frac{(\text{Im} z_{12})^2}{\tau_2}.
\ee
Here the last term is non-holomorphic and comes from the zero mode.
Performing derivatives on above two-point function gives
\be \label{twodp}
\vev{\p_1\phi(z_1,\bar{z}_1)\p_2\phi(z_2,\bar{z}_2)}=-P(z_{12})-\frac{\eta}{w_1}+\frac{\pi}{\tau_2},
\ee
\be \label{ppbar}
\vev{\p_1\phi(z_1,\bar{z}_1)\bar{\p}_2\phi(z_2,\bar{z}_2)}=-\frac{\pi}{\tau_2}.
\ee
The holomorphic and anti-holomorphic stress tensor for boson are $T=-\frac{1}{2}(\p \phi)^2$, $\bar{T}=-\frac{1}{2}(\bar{\p}\phi)^2$ respectively. The expectation value can be calculated by point-splitting
\be \label{T1pt}
\vev{T_{zz}}=-\frac{1}{2}\lim_{z_1\to z_2}\Big(\vev{\p_1\phi(z_1,\bar{z}_1)\p_2\phi(z_2,\bar{z}_2)}+\frac{1}{z_{12}^2}\Big)= \eta-\frac{\pi}{ 2\tau_2},
\ee
where eq.(\ref{twodp}) is used. Note this result is consistent with eq.(\ref{tnox}).
\footnote{which can be verified with the help of the identity for Dedekind $\eta$ function
\be \label{ideneta}
\frac{\p_{\tau} \eta}{\eta}=\frac{i}{2\pi}\eta.
\ee}

Using Wick contraction and eq.(\ref{ppbar}), we can further compute the expectation value
\be\ba
\vev{T(z_1)\bar{T}(z_2)}=&\frac{1}{4}\vev{:(\p\phi(z_1\bar{z}_1))^2::(\bar{\p}\phi(z_2,\bar{z}_2))^2:}\\
=&\frac{1}{2}(\vev{\p_1\phi(z_1,\bar{z}_1)\bar{\p}_2\phi(z_2,\bar{z}_2)})^2+\vev{T_{zz}}\vev{T_{\bar{z}\bar{z}}}\\
=&\eta\bar{\eta}-\frac{\pi\bar{\eta}}{2\tau_2}
-\frac{\pi\eta}{2\tau_2}+\frac{3\pi^2}{4\tau_2^2},
\ea\ee
which is equal to eq.(\ref{vevttb}) as
\be \ba
&\vev{T\bar{T}}=4\pi^2\frac{1}{Z}\p_\tau \p_{\bar{\tau}}Z=\eta\bar{\eta}-\frac{\pi\bar{\eta}}{2\tau_2}
-\frac{\pi\eta}{2\tau_2}+\frac{3\pi^2}{4\tau_2^2}.
\ea\ee
Note that $\vev{T\bar{T}}$ is more complicated than $\vev{TT}$ \cite{Dijkgraaf:1996iy}, since in the latter case, the two holomorphic stress tensor $T$ have more complicated OPE than that of $T$ and $\bar{T}$.

Next we go on to study free fermions case.
The two-point functions for fermion with different spin structure (Denoted by $\n$) are \cite{DiFrancesco:1997nk}
\footnote{ Here the function $P_\n(z)$ is defined by \cite{ellipticbook}
\be\ba \label{pv}
P_\n(v)=\sqrt{P(v)-e_{\n-1}}=\frac{\vartheta_\n(v)\p_z \vartheta_1(0)}{2w_1\vartheta_\n(0) \vartheta_1(v)} ,~~\n=2,3,4.
\ea\ee
}
\be \label{ppv}
\vev{\psi(z)^*\psi(w)}_\n=P_\n(z-w),~~\n=2,3,4.
\ee
The  partition function $Z_\n$ is product of holomorphic and antiholomorphic part
\be
Z_\n=Z'_\n\bar{Z}'_\n,~~Z'_\n(\tau)=\Big(\frac{\vartheta_{\n}(\tau)}{\eta(\tau)}\Big)^{1/2}.
\ee
The holomorphic stress tensor is given by
\be \label{TTpsi}
T=\frac{1}{2}(\p \psi^* \psi-\psi^*\p \psi).
\ee
and similar for the anti-holomorphic part. By subtracting the divergent part, the expectation value is
\be\ba
\vev{T}_\n=&-\frac{1}{2}\lim_{z-w}\Big(\frac{1}{2}(\psi^*(z)\p_w\psi(w)-\p_z \psi^*(z)\psi(w))-\frac{1}{(z-w)^2}\Big)=\frac{1}{4}\frac{\vtv''}{\vtv}-\frac{1}{12}\frac{\vt1'''}{\vt1'},
\ea\ee
which can be shown to be consistent with eq.(\ref{tnox})
on account of the identity $\eta=-\frac{1}{6}\frac{\vt1'''}{\vt1'}$ and eq.(\ref{ideneta})
\be\ba
\vev{T}_\n=2\pi i\p_\tau\ln Z'_\n=i\pi \Big(\frac{\p_\tau\vtv}{\vtv}-\frac{i\eta}{2\pi}\Big)=\frac{1}{4}\frac{\vtv''}{\vtv}-\frac{1}{12}\frac{\vt1'''}{\vt1'}.
\ea\ee
Using wick theorem
\be
\vev{T_{zz}T_{\bar{w}\bar{w}}}_\n= \vev{T}_\n\vev{\bar{T}}_\n=2\pi i\p_\tau\ln Z'_\n \times(-2 \pi i)\p_{\bar{\tau}}\ln \bar{Z}'_\n =4\pi^2\frac{1}{Z_\n} \p_\tau \p_{\bar{\tau}}Z_\n,
\ee
which indicates that eq.(\ref{vevttb}) is valid for free fermions.
\section{Correlation functions of stress tensor}\label{sec:nTTbar}
In this section we will study the correlation functions of stress tensor under $T\bar{T}$ deformed theory up to the first order. This analysis involves multiple $T$'s and $\bar{T}$'s correlation functions in CFTs, which is closely related to the multiple $T$'s correlation functions studied in \cite{Felder:1989vx}. We begin with reviewing how to obtain the correlation functions with multiple $T$'s insertion and then extend to correlation functions with $T$'s and $\bar{T}$'s insertions. For simplicity, we will take  $TT$ inserted correlation function as an example in the following.

We begin with replacing $X$ in eq.(\ref{wipre}) with $T(v)X$, which is
\be\ba \label{ttpre}
&\tr( T(w)T(v) Xq^{L_0-c/12})\\=&2\pi i\frac{\p}{\p \tau}\tr(T(v)Xq^{L_0-c/24})\\+ &  \frac{1}{2\pi i}\oint_{\gamma'} dw_0  \Big(- \zeta(w_0-w)+2\eta_1(w_0-w)-\pi i\Big) \tr( T(w_0)  T(v)Xq^{L_0-c/24}),
\ea\ee
where the contour $\gamma'$ encloses $w_i$ as well as $v$. To perform the contour integral, the following OPE beside eq.(\ref{OPETp}) are needed
\be
T(w)T(v)\sim \frac{c/2}{(w-v)^4}+\frac{2T(v)}{(w-v)^2}+\frac{\p T(v)}{(w-v)}.
\ee
After computing the integral and using translation symmetry, we obtain $TT$ inserted correlation functions \cite{Felder:1989vx}
\be\ba \label{doubleT}
&\tr( T(w)T(v) Xq^{L_0-c/12})\\=&2\pi i\frac{\p}{\p \tau}\tr(T(v)Xq^{L_0-c/24})+\frac{c}{12}P''(v-w) \tr( Xq^{L_0-c/24})\\
&+ 2 \Big(P(w-v)+2\eta_1 \Big)\tr(  T(v) Xq^{L_0-c/24}) + \Big(\zeta(w-v)+2\eta_1 v \Big)\p_v\tr(  T(v) Xq^{L_0-c/24})\\
&+\sum_i
h_i\Big[\Big(P(w-w_i)+2\eta_1 \Big)+\Big( \zeta(w-w_i)+2\eta_1 w_i\Big) \p_{w_i}\Big]\tr(T(v)Xq^{L_0-c/24}).
\ea\ee
With eq.(\ref{doubleT}) in hand, it is readily to write down the expression for multiple $T$'s case
\be\ba \label{nT}
&\tr( T(w)T(v_1)...T(v_n) Xq^{L_0-c/12})\\=&2\pi i\frac{\p}{\p \tau}\tr(T(v_1)...T(v_n)Xq^{L_0-c/24})\\
&+\sum_{j}\frac{c}{12}P''(v-w) \tr(T(v_1)... \hat{T}(v_j)...T(v_n) Xq^{L_0-c/24})\\
&+  2\Big(P(w-v_j)+2\eta_1 \Big)\tr( T(v_1)...T(v_n) Xq^{L_0-c/24})\\
&+\sum_j \Big(\zeta(w-v_j)+2\eta_1 v_j \Big)\p_{v_j}\tr( T(v_1)...T(v_n) Xq^{L_0-c/24})\\
&+\sum_i
h_i \Big(P(w-w_i)+2\eta_1 \Big)\tr(T(v_1)...T(v_n)Xq^{L_0-c/24})  \\
& +  \sum_i \Big( \zeta(w-w_i)+2\eta_1 w_i\Big) \p_{w_i}\tr(T(v_1)...T(v_n) X  q^{L_0-c/24}),
\ea\ee
where a hat on $T$ means the corresponding stress tensor is absent. This is the recursion relations for multiple $T$'s correlation functions \cite{Felder:1989vx}. Next we will consider the  correlation functions with multiple $T$'s and $\bar{T}$'s insertion. For example, adding one $\bar{T}$ to eq.(\ref{ttpre}), one can obtain
\be\ba \label{tttbpre}
&\tr( T(w)T(u)\bar{T}(\bar{v}) Xq^{L_0-c/12})\\=&2\pi i\frac{\p}{\p \tau}\tr(T(u)\bar{T}(\bar{v})Xq^{L_0-c/24})\\&+   \frac{1}{2\pi i}\oint_{\gamma'} dw_0  \Big(- \zeta(w_0-w)+2\eta_1(w_0-w)-\pi i\Big) \tr( T(w_0) T(u) \bar{T}(\bar{v})Xq^{L_0-c/24}),
\ea\ee
where the contour encloses $u,v,w_i$. Again the contour integral around $v$ makes no contribution. Finally, we obtain recursion relations for multiple $T$'s and $\bar{T}$'s {inserted} correlation functions
\be\ba \label{nmTTb}
&\tr( T(w)[T(u_1)...T(u_n)\bar{T}(v_1)...\bar{T}(v_m)] Xq^{L_0-c/12})
\\=&2\pi i\frac{\p}{\p \tau}\tr(T(u_1)...T(u_n)\bar{T}(v_1)...\bar{T}(v_m)Xq^{L_0-c/24})\\
=&\sum_i h_i \Big(- \zeta'(w_i-w)+2\eta_1 \Big)\tr(T(u_1)...T(u_n)\bar{T}(v_1)...\bar{T}(v_m)Xq^{L_0-c/24})
\\& + \sum_i  \Big(- \zeta(w_i-w)+2\eta_1  w_i -2\eta_1 w-\pi i \Big) \p_{w_i}\tr(T(u_1)...T(u_n)\bar{T}(v_1)...\bar{T}(v_m) e^{L_0-c/24}) \\
&+\frac{c}{12}\sum_j P''(u_j-w) \tr( T(u_1)...\hat{T}(u_j)...T(u_n)\bar{T}(v_1)...\bar{T}(v_m) Xq^{L_0-c/24})\\
&+\sum_j 2 \Big(P(w-u_j)+2\eta_1 \Big)\tr( T(u_1)...T(u_n)\bar{T}(v_1)...\bar{T}(v_m)   Xq^{L_0-c/24})\\
&+ \sum_j\Big(\zeta(w-u_j)+2\eta_1 u_j -2\eta_1 w-\pi i \Big)\p_{u_i}\tr(  T(u_1)...T(u_n)\bar{T}(v_1)...\bar{T}(v_m)  Xq^{L_0-c/24}).
\ea\ee
If we replace $T(w)$ with $\bar{T}(w)$ in the first line, then the anti-holomorphic counterpart formula of eq.(\ref{nmTTb}) can also be derived which is expressed {in terms of} anti-holomorphic quantities.

 Let us apply eq.(\ref{nmTTb}) to evaluate three-point function $\vev{\bar{T}(\bar{v}_1)T(u_2)T(u_1)}$
\footnote{We used the following
\be\ba \label{ttx}
&\vev{T(u_2)T(u_1)}=(2\pi i)^2\p_\tau^2 \ln Z+(2\pi i\p_\tau\ln Z)^2+\frac{c}{12}P''(u_1-u_2)+2(P(u_1-u_2)+2\eta)(2\pi i)\p_\tau\ln Z.
\ea\ee One can refer to the appendix \ref{multi-TD} for details.}
\be\ba \label{tbtt}
&\vev{\bar{T}(\bar{v}_1)T(u_2)T(u_1)}\\
=&-2\pi i \p_{\bar{\tau}}\vev{T(u_2)T(u_1)}-2\pi i\vev{T(u_2)T(u_1)}\p_{\bar{\tau}}\ln Z\\
=&\frac{8i\pi^3\p^2_\tau\p_{\bar{\tau}}Z}{Z}+2(P(u_1-u_2)+2\eta)(4\pi^2)\frac{\p_\tau\p_{\bar{\tau}}Z}{Z} +\frac{c}{12}P''(u_1-u_2)(-2\pi i)\p_{\bar{\tau}}\ln Z.
\ea\ee
where the last line does not dependent on $\bar{v}_1$.
With the help of eq.(\ref{tbtt}), we can obtain the four-point function $\vev{\bar{T}(\bar{v}_1)\bar{T}(\bar{v}_2)T(u_2)T(u_1)}$
\be\ba\label{eq60}
&\vev{\bar{T}(\bar{v}_1)\bar{T}(\bar{v}_2)T(u_2)T(u_1)}\\
=&-2\pi i\p_{\bar{\tau}}\vev{T(u_2)T(u_1)\bar{T}(\bar{v}_1)}-2\pi i\vev{T(u_2)T(u_1)\bar{T}(\bar{v}_1)}\p_{\bar{\tau}}\ln Z\\&+\frac{c}{12}\bar{P}''(\bar{v}_{12})\vev{T(u_1)T(u_2)}+2(-\bar{\zeta}'(\bar{v}_{12})+2\bar{\eta})\vev{T(u_2)T(u_1)\bar{T}(\bar{v}_1)}\\&
+(-\bar{\zeta}(\bar{v}_{12})+2\bar{\eta}\bar{v}_{12}+\pi i)\p_{\bar{v}_1}\vev{T(u_2)T(u_1)\bar{T}(\bar{v}_1)},
\ea\ee
where $\bar{v}_{12}=\bar{v}_1-\bar{v}_2$. Note the last term equals zero since $\vev{T(u_2)T(u_1)\bar{T}(\bar{v}_1)}$ is independent of $\bar{v}_1$. Finally, the eq.(\ref{eq60}) can be expressed as
\be\ba \label{Fourt}
 &\vev{\bar{T}(\bar{v}_1)\bar{T}(\bar{v}_2)T(u_2)T(u_1)}\\
=&\frac{16\pi^4}{Z}\p^2_\tau\p^2_{\bar{\tau}}Z+(2\pi i)^2\frac{c}{12}(P''(u_{12})\p^2_{\bar{\tau}}\ln Z+\bar{P}''(\bar{v}_{12})\p^2_\tau\ln Z)\\
&+2(2\pi i)^3\frac{1}{Z}((P(u_{12})+2\eta)\p_\tau\p^2_{\bar{\tau}}Z-(\bar{P}(\bar{v}_{12})+2\bar{\eta})\p^2_\tau\p_{\bar{\tau}}Z)\\
&+(2\pi i)^2\frac{c}{12}(P''(u_{12})(\p_{\bar{\tau}}\ln Z)^2+\bar{P}''(\bar{v}_{12})(\p_\tau\ln Z)^2)
\\&+\frac{c}{12}4\pi i(\bar{P}''(\bar{v}_{12})(P(u_{12})+2\eta)\p_\tau\ln Z-P''(u_{12})(\bar{P}(\bar{v}_{12})+2\bar{\eta})\p_{\bar{\tau}}\ln Z)
\\
&+\Big(\frac{c}{12}\Big)^2\bar{P}''(\bar{v}_{12})P''(u_{12})+4(2\pi)^2(\bar{P}(\bar{v}_{12})+2\bar{\eta})(P(u_{12})+2\eta)\frac{1}{Z}\p_\tau\p_{\bar{\tau}}Z.
\ea\ee

To obtain the first order deformed correlation function, one has to do the integral eq.(\ref{eq1}) on the torus. To illustrate how to construct the first order correction from the correlation functions in CFTs, we take the eq.(\ref{tbtt}) and eq.(\ref{Fourt}) as two examples.

Firstly, from eq.(\ref{tbtt}), we can compute the deformed one-point function $\vev{T}_\l$ up to the first order
\be\ba
\vev{T(u_1)}_\l=&\frac{\int D\phi T(u_1)e^{-S_0+\l\int d^2uT\bar{T}(u)}}{\int D\phi  e^{-S_0+\l\int d^2uT\bar{T}(u)}}\\
=&\vev{T} -\l\vev{T} \int d^2u\vev{T\bar{T}(u)}+\l \int d^2u\vev{T\bar{T}(u)T(u_1)}+...
\ea\ee
Here $S_0$ is the action of CFT, the correlation function $\vev{...}$ is evaluated in the un-deformed theory. The integral in the second term comes from the correction of partition function, which is considered in previous section. This term is vanishing on plane. The integrand in the last term can be obtained by setting $v_1=u_1$ in  eq.(\ref{tbtt}). Finally, by performing the integral explicitly, we obtain \footnote{Please refer to appendix \ref{intfor} for details.}
\be \label{int1p}
\vev{T}_\l-\vev{T}=\l\Big(\frac{(2\pi i)^3\tau_2\p^2_{\tau}\p_{\bar{\tau}}Z}{Z}-(2\pi)^3\frac{\p_\tau\p_{\bar{\tau}}Z}{Z}
+(2\pi i)\frac{\p_\tau Z}{Z}\frac{(2\pi)^2\tau_2\p_\tau\p_{\bar{\tau}}Z}{Z}\Big).
\ee
In computing these integrals, following the prescription {for} regularization in  \cite{Dijkgraaf:1996iy}, we have removed the singular points out the integration domain.

Secondly, one can consider the two-point function $\vev{T\bar{T}}_\l$ up to the first order as follows
\be\ba
\vev{T(u_2)\bar{T}(\bar{v}_2)}_\l=&\vev{T \bar{T} }-\l\vev{T \bar{T} }\int d^2u\vev{T\bar{T}(u)}+\l \int d^2u\vev{T\bar{T}(u)T(u_2)\bar{T}(\bar{v}_2)}...
\ea\ee
where only the last term is unknown. The integrand in the last term can be obtained by substituting eq.(\ref{Fourt}) with $v_1=u_1$. It turns out the last term is
\be \label{TTbTTb}
\int d^2u\vev{T\bar{T}(u)T(u_2)\bar{T}(\bar{v}_2)}=\frac{16\pi^4}{Z}(\tau_2\p^2_\tau\p^2_{\bar{\tau}}Z  -i( \p_\tau\p^2_{\bar{\tau}}Z-\p^2_\tau\p_{\bar{\tau}}Z)),
\ee
where the detailed calculation  is presented in appendix \ref{intfor}.

In addition, we can also calculate the correlation $\vev{TT}_\l$  up to the first order
\be\ba \label{T4}
\vev{T(u_1T(u_2)}_\l=&\vev{T(u_1)T(u_2)}-\l\vev{T(u_1)T(u_2)}\int d^2u\vev{T\bar{T}(u)}\\
&+\l \int d^2u\vev{T\bar{T}(u)T(u_1)T(u_2)},
\ea\ee
where the integral in the last line is more involved than (\ref{TTbTTb}) and the computation details are presented in appendix \ref{deTT}. The final result turns out to be
\be \ba
&\int  d^2u\vev{T\bar{T}(u)T(u_1)T(u_2)}\\
=&2\pi i \Big( \frac{3(2\pi )^3 \p^2_\tau\p_{\bar{\tau}}Z}{2Z}-\frac{(2\pi i)^3\tau_2\p^3_\tau\p_{\bar{\tau}}Z}{Z}\Big)+\frac{c\tau_2}{12}P''(u_1-u_2)\vev{T\bar{T}} \\
&+ \frac{16i\pi^4\p^2_\tau\p_{\bar{\tau}}Z}{Z} +\frac{(16\pi^2)\p_\tau\p_{\bar{\tau}}Z}{Z}( P_{u_1,u_2} +2\eta \pi )
 \\&+2(P(u_1-u_2)+2\eta )\Big(-\frac{(2\pi i)^3\tau_2\p^2_\tau\p_{\bar{\tau}}Z}{Z}+(2\pi )^3\frac{\p_\tau\p_{\bar{\tau}}Z}{Z}\Big) \\
 &+ (8\pi^2)\frac{\p_\tau\p_{\bar{\tau}}Z}{Z}\Big(-2i\eta\eta' -P_{u_1,u_2}+2i \bar{\tau}\eta^2   -2\pi\eta \Big).
\ea\ee
where $P_{a,b}$ is defined and calculated in (\ref{pab}).
\section{Deformed correlation functions in path integral formalism}\label{sec:TTbarPI}
In this section, we will derive the correlation functions with $T\bar{T}$ insertion in CFT defined on torus, following the line of \cite{Eguchi:1986sb} where the $TT$ insertion was obtained in path integral formalism. We start with the definition of stress tensor, assuming there is a Lagrangian description for the theory
\be\ba
T_{\m\n}=-\frac{2}{\sqrt{g}}\frac{\p S}{\p g^{\m\n}},
\ea\ee
where $S$ is the CFT action, then the expectation value of stress tensor is given by
\be
\vev{T_{\m\n}}=\frac{2}{Z\sqrt{g}}\frac{\delta}{\delta g^{\m\n}}Z
,~~
Z=\int D\phi e^{-S}.
\ee
The correlation functions are defined by
\be
\vev{X}=\frac{1}{Z}\int d\phi Xe^{-S},~X=\phi_1...\phi_N.
\ee
The Ward identity corresponding to three types of local symmetries: reparametrization, local rotation and Weyl scaling in CFT can be written as \cite{Eguchi:1986sb}
\be\ba \label{WI87}
&\frac{1}{2}\int d^2x\sqrt{g}e^a_\n (P\xi)^{\n\m}\vev{T^a_\m(x)X}\\=& -\sum_{k=1}^N\Big(\xi^\m(x_k)\p^k_\m+\frac{d_k}{2}\nabla_\rho\xi^\rho+i s_k\Big(\frac{1}{2}\epsilon_{\rho\sigma}\nabla^\rho\xi^\sigma+\omega_\n\xi^\n\Big)\Big)\vev{X}+\frac{c}{48\pi}\int d^2x\sqrt{g}R\nabla_\rho \xi^\rho \vev{X}.
\ea\ee
where $e^a_\m$ is the zweibein field coupled with CFT and $\omega_\n$ is the spin connection. The vector fields $\xi^\m$  parameterize  the transformation of zweibein: $e^\m_a\to e^\m_a-\xi^\n\p_\n e^\m_a+\p_\n\xi^\m e^\n_a$. $s_k,d_k$ are the spin and dimension of the field $\phi_k$ respectively. $R$ is the scalar curvature of the surface, which is equal to zero for torus.  And
\be
(P\xi)^{\n\m}=G_{\rho\s}^{~~~\n\m}\nabla^\rho\xi^\s,~~G_{\rho\s}^{~~~\n\m}=\delta_\rho^\n\delta_\s^\m+\delta_\rho^\m\delta_\s^\n-g^{\m\n}g_{\rho\s}.
\ee
In order to obtain double stress tensors insertions, one can further vary eq.(\ref{WI87}) with respect to metric. The resulting expression is
\be \ba \label{varWI}
&\frac{1}{4}  (G_{\rho\s}^{~~~\m\l}\nabla_\l\xi^\n+G_{\rho\s}^{~~~\n\l}\nabla_\l\xi^\m+ G_{\rho\s}^{~~~\m\n}\xi^\l\nabla_\l)\vev{T_{\m\n}(w)X}\\
&+\frac{1}{4}\int d^2z (\sqrt{g} (P\xi)^{\m\n}) \vev{T_{\m\n}(z)T_{\rho\s}(w)X}\\
=&-\frac{1}{2}\sum_k\Big(\xi^\m(x_k)\p^k_\m+\frac{d_k}{2}\nabla_\a\xi^\a+i s_k\Big(\frac{1}{2}\epsilon_{\a\beta}\nabla^\a\xi^\beta+\omega_\n\xi^\n\Big)\Big)\vev{T_{\rho\s}(w)X} \\
&+\frac{c}{96\pi }(-2\nabla_{(\rho}\nabla_{\s)}\nabla_\l \xi^\l+2g_{\rho\s}\nabla^2\nabla_\l\xi^\l+\nabla_\l(R\xi^\l) g_{\rho\s})\vev{X}\\
&+\frac{c}{96\pi }\int d^2z\sqrt{g}R\nabla_\l\xi^\l\vev{T_{\rho\s}(w)X}.
\ea\ee
By setting $\rho=\s=z$ and $\xi^{\bar{z}}=0$ in eq.(\ref{varWI}),  the correlation functions with $TT$ insertion can be obtained as presented in \cite{Eguchi:1986sb}. Similarly, the $T\bar{T}$ insertion can be obtained by setting $\rho=\s=\bar{z}$ and $\xi^{\bar{z}}=0$. Then eq.(\ref{varWI}) turns out to be
\be \ba \label{TTb}
&  \frac{1}{2}\int d^2z  \sqrt{g} (P\xi)^{zz}  \vev{T_{zz}(z)T_{\bar{w}\bar{w}}(w)X}+\xi^w\nabla_w \vev{T_{\bar{w}\bar{w}}(w)X})\\
=&-\sum_k\Big( h_k\nabla_{w_k}\xi^{w_k}+\xi^{w_k}(\p_{w_k}+is_k\omega_{w_k})\Big)\vev{T_{\bar{w}\bar{w}}(w)X}\\
& +\frac{c}{24\pi }(-\nabla_{\bar{w}}\nabla_{\bar{w}}\nabla_w \xi^w )\vev{X}+\frac{c}{48\pi }\int d^2z\sqrt{g}R\nabla_z\xi^z\vev{T_{\bar{w}\bar{w}}(w)X},
\ea\ee
where $h_k=\frac{1}{2}(d_k+s_k)$ and we omitted the term $\vev{T_{\bar{z}z}...}$. To extract the $\vev{T_{zz}(z)T_{\bar{w}\bar{w}}(w)X}$ outside the integral on the RHS of eq.(\ref{TTb}), the Green function $G^z_{~vv}$ for operator $\nabla^z$  on Riemann surface with genus $g$ is employed \cite{Eguchi:1986sb}
\be\ba
\nabla^z G^z_{~vv}(z,v)=\frac{1}{\sqrt{g}}\delta^{(2)}(z-v)-\sum_{j=1}^{3g-3}g^{z\bar{z}}\eta^z_{~\bar{z},j}(z,\bar{z})h_{vv}^{~~j}(v),
\ea\ee
where $h_{vv}^{~~j}(v)$ are holomorphic quadratic differentials on the Riemann surface, and $\eta^z_{~\bar{z},i}$ are Beltrami differentials dual to holomorphic quadratic differentials, i.e., $\int d^2z \sqrt{g}g^{z\bar{z}}h_{zz}^{~~j}\eta^z_{~\bar{z},i}=\delta_i^j$.
Let $\xi^z(z)=G^z_{~vv}(z,v)$, then eq.(\ref{TTb}) can be written as
\be \ba \label{TTbar1}
& \vev{T_{vv}(v )T_{\bar{w}\bar{w}}(w)X}-\sum_j h_{vv}^{~~j}(v)\int d^2z\sqrt{g}g^{z\bar{z}}(z)\eta^z_{~\bar{z},j}(z)\vev{T_{zz}(z)T_{\bar{w}\bar{w}}(w)X}\\=&-G^w_{~vv}(w,v)\nabla_w \vev{T_{\bar{w}\bar{w}}(w)X})\\
&-\sum_k\Big( h_k\nabla_{w_k}G^{w_k}_{~~vv}(w_k,v)+G^{w_k}_{~~vv}(w_k,v)(\p_{w_k}+is_k\omega_{w_k})\Big)\vev{T_{\bar{w}\bar{w}}(w)X}\\
& -\frac{c}{24\pi }( \nabla_{\bar{w}}\nabla_{\bar{w}}\nabla_w G^w_{~vv}(w,v) )\vev{X}+\frac{c}{48\pi }\int d^2z\sqrt{g}R\nabla_zG^z_{~vv}\vev{T_{\bar{w}\bar{w}}(w)X},
\ea\ee
where the last term  on LHS is called Teichmuller term.
All the formulae derived so far are valid for general Riemann surface. Here we are interested in the case $g=1$, i.e., the torus, in which case the metric are flat ($R=0$), $y^j=-\tau$, and the corresponding Beltrami differential and quadratic differential for torus are
\be
\eta^z_{~\bar{z}}=\frac{i}{\text{Im}\tau},~~h_{zz}=-i.
\ee
The explicit expressions for $  G^z_{~vv}(z,v)$ on torus is
\be
G^z_{ww}(z,w)=\frac{1}{2\pi}\frac{\vartheta'_1(z-w)}{\vartheta_1(z-w)}+i\frac{\text{Im}(z-w)}{\text{Im}\tau}.
\ee
With these parameters in hand, the Teichmuller term  can be computed explicitly as
\be\ba \label{tei1}
&h_{zz}^j(z )\int d^2v\sqrt{g}g^{v\bar{v}}(v)\eta^v_{~\bar{v},j}(v)\vev{T_{vv}(v)T_{\bar{w}\bar{w}}(w)X} \\
=&\oint dz\vev{T_{zz}(z)T_{\bar{w}\bar{w}}(w)X}+2i\int d^2z\sqrt{g}\frac{\text{Im}z}{\text{Im}\tau}\p_{\bar{z}}\vev{T_{zz}(z)T_{\bar{w}\bar{w}}(w)X},
\ea\ee
where the last term can be evaluated by substituting eq.(\ref{TTbar1}).  The derivative in the last term does not vanish, since the correlation function can be non-analytical in $z$ as $T_{zz}(z)$ approaches other operators.
As for the first term, it turns out to be
\footnote{
In this section, in order to compare our results to that of \cite{Eguchi:1986sb}, we follow the convention in that paper, where the stress tensor on torus is related to previous section upto a factor $2\pi$, and the stress tensor on plane $T_{pl}$ is the same with previous definition, thus eq.(\ref{TransT}) become
\be
w^{\prime2} T_{pl}(w')=\frac{2\pi}{(2\pi i)^2}T(w)+\frac{c}{24},~~~w'=e^{2\pi i w}.
\ee
Here $T_{pl}(w')=\sum L_n/w^{\prime n+2},T(w)=(-2\pi)\sum e^{-2\pi i w n}(L_{cy})_n$, with $(L_{cy})_n=L_n-\delta_{n,0}c/24$, then
\be \ba
&\oint dw\vev{T_{ww}(w)T_{\bar{v}\bar{v}}(v)X}=-2\pi \vev{(L_{cy})_0T_{\bar{v}\bar{v}}(v) X}=-\frac{1}{Z}q\frac{\p}{\p q}\tr(q^{(L_{cy})_0}T_{\bar{v}\bar{v}}(v)X)\\=&i\p_\tau\vev{T_{\bar{v}\bar{v}}(v)X}+i\p_\tau \ln Z\vev{T_{\bar{v}\bar{v}}(v)X}.
\ea\ee
}
\be\ba
\oint dz\vev{T_{zz}(z)T_{\bar{w}\bar{w}}(w)X}=i\p_\tau\vev{ T_{\bar{w}\bar{w}}(w)X}+i\p_\tau\ln Z\vev{T_{\bar{w}\bar{w}}(w)X}.
\ea\ee
Finally the Teichmuller term is
 \be\ba \label{tei1}
&h_{zz}^j(z )\int d^2v\sqrt{g}g^{v\bar{v}}(v)\eta^v_{~\bar{v},j}(v)\vev{T_{vv}(v)T_{\bar{w}\bar{w}}(w)X} \\
 =&i\p_\tau\vev{ T_{\bar{w}\bar{w}}(w)X}+i\p_\tau\ln Z\vev{T_{\bar{w}\bar{w}}(w)X}+\Big(i\frac{\text{Im}w}{\text{Im}\tau} \Big)\p_w\vev{T_{\bar{w}\bar{w}}(w)X}\\
&+\frac{1}{2}\sum_k h_k\frac{1}{\text{Im}\tau}\vev{T_{\bar{w}\bar{w}}(w)X}+i\sum_k\frac{\text{Im}w_k}{\text{Im}\tau}\p_{w_k}\vev{T_{\bar{w}\bar{w}}(w)X}.
\ea\ee

Combining with the remaining terms in eq.(\ref{TTbar1}) which can be computed straightforwardly, the $T\bar{T}$ inserted correlation function is given by
\be\ba \label{PIWard}
&\vev{T_{zz}(z)T_{\bar{w}\bar{w}}(w)X}\\
=&i\p_\tau\vev{ T_{\bar{w}\bar{w}}(w)X}+i\p_\tau\ln Z\vev{T_{\bar{w}\bar{w}}(w)X} \\
&-\sum_k \Big(h_k(\frac{1}{2\pi}(\xi'(w_k-z)-2\eta_1))+(\frac{1}{2\pi}(\xi(w_k-z)-2\eta_1(w_k-z)))\p_{w_k}\Big)\vev{T_{\bar{w}\bar{w}}(w)X}\\
&-\Big(\frac{1}{2\pi}(\xi(w-z)-2\eta_1(w-z))\Big)\p_w\vev{T_{\bar{w}\bar{w}}(w)X}-\frac{c}{48\pi} \p_{\bar{w}} \p_w \delta(w-z)\vev{X},
\ea\ee
where the term $\p_w\vev{T_{\bar{w}\bar{w}}(w)X}$  in last line does not vanish since $\vev{T_{\bar{w}\bar{w}}(w)X}$ is not analytic in $w$ as $T_{\bar{w}\bar{w}}$ goes to $X$, as mentioned before. In fact, $\p_w\vev{T_{\bar{w}\bar{w}}(w)X}$ is proportional to delta functions such as $\delta^{(2)}(w-w_k)$ (which can be seen by substituting the expression of one $\bar{T}$ inserted function $\vev{T_{\bar{w}\bar{w}}X}$).  Therefore the terms in the last line of eq.(\ref{PIWard}) are contact terms. In addition, the term $\sum_k z\p_{w_k}\vev{T_{\bar{w}\bar{w}(w)}X}$ is also contact term (see eq.(\ref{cont})). As discussed around eq.(\ref{firord}), when we consider the first order of $T\bar{T}$ deformed correlation functions, the contact points is dropped out from the integral.
Upon ignoring the contact terms eq.(\ref{PIWard}) is consistent with the result in  section \ref{sec:TTbar}. Therefore the operator formalism and path integral method are consistent with each other when we consider the first order $T\bar{T}$ deformed correlation functions.
\section{Conclusions and discussions}
Motivated by studying the quantum chaos, the quantum entanglement of the local excited states in $T\bar{T}$ field theories, one has to know the correlation {functions} on torus with {the $T\bar{T}$ deformation}. In this work, to construct the correlation functions of the CFTs on torus with $T\bar{T}$ deformation, we apply the Ward identity on torus and do a proper regularization procedure to figure out the correlation functions with $T\bar{T}$ deformation in terms of perturbative field theory approach. It can be regarded as a direct generalization of previous studies \cite{He:2019vzf}\cite{He:2019ahx} on correlation functions in the $T\bar{T}$ deformed bosonic and supersymmetric CFTs defined on plane. It is well known that the the correlation functions on plane with $T$ and $\bar{T}$ can be obtained straightforwardly by using the Ward identity, while the Ward identity on the torus is very complicated and Ward identity associated with the $T$ and $\bar{T}$ is unknown in the literature. In this work, we obtained the $T\bar{T}$ deformed correlation functions perturbatively in both operator formalism and in path integral language. As a consistent check, the first order correction to the partition function agrees with that obtained by different approach \cite{Datta:2018thy} in literature. We explicitly calculate the first order correction to partition function in the free field theories and  we confirm the validity by comparing with the results obtained by Wick contraction.  Moreover, we obtain a new recursion relations of the correlation functions of the multiple $T$'s and $\bar{T}$'s insertion in generic CFTs on torus, with which we also figure out the some closed form of the first order $T\bar{T}$ corrections to the correlation functions of stress tensors.

Since resulting correlation functions are applicable for generic CFTs with the deformation, they are useful to study the holographic aspects of the dual boundary CFTs with finite size, finite temperature effects. In addition, it is interesting to investigate the correlation functions of  the supersymmetric theories on the torus, as we did in \cite{He:2019ahx}.

\subsection*{Acknowledgements}
We would like to thank Bin Chen, Hao Geng, Yongchao Lv, Hongfei Shu, Jia-Rui Sun and Stefan Theisen for useful discussion. S.H. would like to appreciate the financial support from Jilin University and Max Planck Partner group.
Y.S. would like to thank to the support from China Postdoctoral Science Foundation (No. 2019M653137).
\appendix
 \appendixpage
 \addappheadtotoc
\section{Conventions}\label{conventions}
In our convention the torus denoted as $T^2$ is defined by the identification of complex number $w\sim w+2w_1+2w_2$ with $2w_1=1,2w_2=\tau$.

In the following we collect some formulae regarding elliptic functions which are useful in this work.
The Weierstrass $P$-function is defined by \cite{ellipticbook}
\be
P(z)=\frac{1}{z^2}+\sum_{n,m\neq 0}\Big(\frac{1}{(z-\omega_{n,m})^2}-\frac{1}{\omega_{n,m}^2}\Big),~~\omega_{n,m}=2w_1n+2w_2m.
\ee
The Weierstrass $P$-function is an elliptic function (doubly periodic on complex plane) with periods $2w_1$ and $2w_2$.  $P(z)$ is even and has only one second order pole at $z=0$ on torus. The Laurent series expansion in the neighborhood of $z=0$ can be expressed as
\be
P(z)=\frac{1}{z^2}+c_2 z^2+c_4 z^4+...
\ee
where $c_{2n}$ are constants.

The Weierstrass $\zeta(z)$ function is defined by
\be
\zeta(z)=\frac{1}{z}+\sum_{n,m\neq 0}\Big(\frac{1}{z-\omega_{n,m}}+\frac{1}{\omega_{n,m}}+\frac{z}{\omega_{n,m}^2}\Big),~~\omega_{n,m}=2w_1n+2w_2m,
\ee
which is related with $P(z)$ as
\be
P(z)=-\zeta'(z).
\ee
Note $\zeta(z)$ is odd and has a simple pole at $z=0$
around which the Laurent expansion takes the form
\be
\zeta(z)=\frac{1}{z}-\frac{c_2}{3}z^3-\frac{c_4}{5}z^5+...
\ee
Since an elliptic function {can} not have only one simple pole on torus, $\zeta(z)$ is not doubly periodic. Instead, $\zeta(z)$ satisfies the quasi-doubly periodic conditions
\be \label{quasip}
\zeta(z+2w_{1,2})=\zeta(z)+2\zeta(w_{1,2})
\ee
with $\zeta(w_1)$ equals the Dedekind $\eta$ function (also denoting $\eta_1\equiv \zeta(w_1)$) and  $\zeta(w_2)\equiv\eta'$. These quantities satisfy the following identity
\be \label{Identity}
\eta w_2-\eta' w_1=\frac{\pi i}{2}.
\ee
From $\zeta(z)$ function, the $\sigma(z)$  function is defined as
\be
\zeta(z)=\p_z\ln \s(z).
\ee
The $\s(z)$ function has the following properties
\be
\s(z+2w_1)=-e^{2\eta(z+w_1)}\s(z),~~\s(z+2w_2)=-e^{2\eta'(z+w_2)}\s(z).
\ee
\section{Useful integrals}\label{intfor}
In this section, the Stoke's theorem in 2D is frequently used and it is
\be
\int_M dz\wedge d\bar{z}(\p_z F^z+\p_{\bar{z}}F^{\bar{z}})=\oint_{\p M} (F^z d\bar{z}-F^{\bar{z}}dz)
\ee
with $dz\wedge d\bar{z}=-2idx\wedge dy=-2i d^2z$. The area of torus $T^2$ is $\int_{T^2}d^2z=\tau_2$, where the torus  is the parallelogram on plane enclosed by $OABC$ with $O:z_0$, $A: z_0+2w_1$, $B: z_0+2w_1+2w_2$, $C:z_0+2w_2$.

In the following we will evaluate the integrals in (\ref{int1p}) which involve the integrals of $P(x-y)$ and $P''(x-y)$ over torus with coordinates $x$. Note both of the functions are singular at $x=y$. To deal with this singularity in the integral, we follow the prescription  in \cite{Dijkgraaf:1996iy} (see also \cite{Douglas:1993wy}) where we cut the singular point out of the integration domain, more precisely, we perform the integral as follows
\be\ba \label{intPW1}
&\int_{T^2-D(y)} d^2z P(z-y)=-\int d^2z\p_z\zeta(z-y) =-\frac{
i}{2}\oint_{\p T^2}d\bar{z}\zeta(z-y)\\
=&-\frac{i}{2} \Big(\int^A_O-\int^B_C\Big)d\bar{z}\zeta(z-y)-\frac{i}{2}\Big(\int_A^B-\int^C_O\Big)d\bar{z}\zeta(z-y)\\
=&-\frac{i}{2}\int_{ 0}^{2w_1}d\bar{z}(\zeta(z-y)-\zeta(z-y+2w_2)) -\frac{i}{2}\int_{0}^{ 2w_2}d\bar{z}(\zeta(z-y+2w_1)-\zeta(z-y))\\
=&-i \bar{w}_1(-2\eta')-i \bar{w}_22\eta= \pi -
4\eta  \text{Im}w_2=  \pi-2 \eta \tau_2,
\ea\ee
where $D(y)$ is a infinitesimal small disk around the singular point. In the last step eq.(\ref{Identity}) is used to eliminate $\eta'$.
One has to be careful when evaluate this integral, since the boundary of integration domain is $\p T^2-\p D(y)$, we must compute the contour integral along the small circle $\p D(y)$. Actually, one can check that the integral above along the contour $\p D(y)$ is zero, making no contribution to the final answer\footnote{Interestingly, it can be checked that in all the integrals considered in this work, if $z_i$ is a singular point of the integrand, the path integrals along  $\p D(z_i)$ vanish. Thus we will not mention the integrals along this kind of path hereafter.}. So we does not write is explicitly out in eq.(\ref{intPW1}). In a similar manners, we can handle the integral $\int d^2z P''(z-y)$ which turns out to be zero. Note the two integrals are exactly equal to the results obtained by using the formalism in \cite{Dijkgraaf:1996iy}.

Next we turn to the integral, similar with eq.(\ref{intPW1}),
\footnote{
In the second step we used the integration by parts. One may worry about that we omit the term
\be\ba
\int d^2u \zeta(u-a)\p_u\bar{P}(\bar{u})= \int d^2u \zeta(u-a)\p_{\bar{u}}\delta^{(2)}(u)=-\int d^2\delta^{(2)}(u)\p_{\bar{u}}\zeta(u-a)=\int d^2u\delta^{(2)}(u)\delta^{(2)}(u-a),
\ea\ee which is divergent as $a=0$. However, this will not cause problem, since the domain of integral does not include the small disk around the singular points $u=a$ and $u=0$, this term will not appear.
}
\be\ba \label{intPP}
&\int_{T^2-D(0)-D(a)} d^2 u P(u-a)\bar{P}(\bar{u})\\=&\int d^2 u (-\zeta'(u-a)\bar{P}(\bar{u}))
=\int d^2 u\p_u(-\zeta(u-a)\bar{P}(\bar{u}))=\frac{i}{2}\oint d\bar{u}(-\zeta(u-a)\bar{P}(\bar{u}))\\
=&\frac{i}{2}\int_{z_0}^{z_0+2w_1}d\bar{u}[-\zeta(u-a)\bar{P}(\bar{u})+\zeta(u-a+2w_2)\bar{P}(\bar{u}+2\bar{w}_2)]\\
&+\frac{i}{2}\int_{z_0}^{z_0+2w_2}d\bar{u}[-\zeta(u-a+2w_1)\bar{P}(\bar{u}+2\bar{w}_1)+\zeta(u-a)\bar{P}(\bar{u})]=2i(\eta\bar{\eta}'-\eta'\bar{\eta}),
\ea\ee
where we used eq.(\ref{quasip}), $P(u)=-\zeta'(u)$ and {the fact} $P(u)$ being doubly periodic function.
It follows that
\be\ba
&\int d^2z(P(z-a)+2\eta)(\bar{P}(z)+2\bar{\eta})=2i(\eta\bar{\eta}'-\eta'\bar{\eta})+2\pi(\eta+\bar{\eta})-4\eta\bar{\eta}\tau_2=0,
\ea\ee
where eq.(\ref{Identity}) is used in the last step.

By the same reason, one has
\be\ba \label{intPP2}
\int d^2z P''(z-a)\bar{P}(\bar{z})=\int d^2z P''(z-a)\bar{P}'(\bar{z})=\int d^2z P''(z-a)\bar{P}''(\bar{z})=0,
\ea\ee
where for example we can write $P''(z)\bar{P}(\bar{z})=\p_z(P'(z)\bar{P}(\bar{z}))$ inside the integral. Note here the integral domain is $T^2-D(0)-D(a)$ as mentioned before.
\section{Details on $\vev{T(u_1)T(u_2)\bar{T}(\bar{v}_1)}$}\label{multi-TD}
In this section we will compute three-point function $\vev{T(u_1)T(u_2)\bar{T}(\bar{v}_1)}$.
We begin {with} introducing several useful formulae obtained by taking derivatives on eq.(\ref{for1})
\be \ba\label{PPP}
(2\pi i)^2\sum_{n\neq 0}\frac{n}{1-q^n}\Big(\frac{z_1}{z_2}\Big)^n=&P(w_1-w_2)+2\eta_1,\\
(2\pi i)^3\sum_{n\neq 0}\frac{n^2}{1-q^n}\Big(\frac{z_1}{z_2}\Big)^n=&P'(w_1-w_2),\\
(2\pi i)^4\sum_{n\neq 0}\frac{n^3}{1-q^n}\Big(\frac{z_1}{z_2}\Big)^n=&P''(w_1-w_2)
\ea\ee
with $z_{1,2}=e^{2\pi iw_{1,2}}$.
We can now evaluate the following trace
\be\ba \label{TpTp}
\tr(q^{L_0-c/24}T_{pl}(z_1)T_{pl}(z_2))=\sum_{n,m}z_1^{-n-2}z_2^{-m-2}\tr(q^{L_0-c/24}L_nL_m),
\ea\ee
where for the term with $n=m=0$, $\tr(q^{L_0-c/24}L_0L_0)$ can be expressed as derivatives of partition function $Z=\tr(q^{L_0-c/24})$ with respect to $\tau$. While for the remaining terms, using eq.(\ref{formula}), we get
\be\ba
 \tr(q^{L_0-c/24}L_nL_m)=q^{-n} \tr(q^{L_0-c/24}L_mL_n),
\ea\ee
which leads to
\be\ba
\tr(q^{L_0-c/24}L_mL_n)=\frac{1}{q^{-n}-1} \tr(q^{L_0-c/24}[L_n,L_m]).
\ea\ee
With the help of Virosoro algebra and eq.(\ref{formula}), we obtain
\be \ba
 \tr(q^{L_0-c/24}L_mL_n)=&\frac{1}{q^{-n}-1}\tr\Big(q^{L_0-c/24}\Big((n-m)L_{n+m}+\frac{c}{12}n(n^2-1)\delta_{m+n,0}\Big)\Big)\\
=&\frac{\delta_{m+n,0}}{q^{-n}-1}\tr\Big(q^{L_0-c/24}\Big(2nL_{0}+\frac{c}{12}n(n^2-1)\Big)\Big).
\ea\ee
Substituting into eq.(\ref{TpTp}), then the summation in eq.(\ref{TpTp}) can be obtained via eq.(\ref{PPP}). With transforming the stress tensor on plane into cylinder, we finally obtain $\vev{T(u_1)T(u_2)}$ in eq.(\ref{ttx}).

To calculate the three-point function $\vev{T(u_1)T(u_2)\bar{T}(\bar{v}_1)}$, one can start with
\be\ba \label{TpTpTp}
\tr(q^{L_0-c/24}T_{pl}(z_1)T_{pl}(z_2)\bar{T}_{pl}(\bar{y}_1))=\sum_{n,m}z_1^{-n-2}z_2^{-m-2}\bar{y}_1^{-r-2}\tr(q^{L_0-c/24}L_nL_m\bar{L}_r),
\ea\ee
where the only non-vanishing trace in the summation is $\tr(q^{L_0-c/24}L_0L_0\bar{L}_0)$ and
\be \ba
\tr(q^{L_0-c/24}L_mL_n\bar{L}_0)= \frac{\delta_{m+n,0}}{q^{-n}-1}\tr\Big(q^{L_0-c/24}\Big(2nL_{0}+\frac{c}{12}n(n^2-1)\Big)\bar{L}_0\Big).
\ea\ee
Following the steps deriving $\vev{T(u_1)T(u_2)}$, we will finally obtain the same express as presented in eq.(\ref{tbtt}). Similarly, the deriving of four-point function $\vev{T(u_1)T(u_2)\bar{T}(\bar{v}_1)\bar{T}(\bar{v}_2)}$ in eq.(\ref{Fourt}) can be proceeded.

\section{Details on $\vev{TT}_\l$}\label{deTT}
In this section we will compute the last integral in eq.(\ref{T4}).
From the recursion relation for $T,\bar{T}$ inserted correlation functions in section \ref{sec:nTTbar}, the four point function interested here takes the form
\be\ba
&\vev{T(w)T(u_1)T(u_2)\bar{T}(\bar{v}_1)}\\
=&2\pi i\p_{\tau}\vev{T(u_1)T(u_2)\bar{T}(\bar{v}_1)}+\vev{T}\vev{T(u_1)T(u_2)\bar{T}(\bar{v}_1)}+\frac{c}{12}(P''(w-u_1)+P''(w-u_2))\vev{T\bar{T}}\\
&+2(P(w-u_1)+2\eta)\vev{T(u_1)T(u_2)\bar{T}(\bar{v}_1)}+2(P(w-u_2)+2\eta)\vev{T(u_1)T(u_2)\bar{T}(\bar{v}_1)}\\
&+(\zeta(w-u_1)+2\eta u_1)\p_{u_1}\vev{T(u_1)T(u_2)\bar{T}(\bar{v}_1)}+(\zeta(w-u_2)+2\eta u_2)\p_{u_2}\vev{T(u_1)T(u_2)\bar{T}(\bar{v}_1)}.
\ea\ee
Letting $u_1=v_1$ and integrating $u_1$ over torus, we obtain the last integral in eq.(\ref{T4})
\be\ba  \label{TTTTar}
&\int  d^2u_1\vev{T(w)T(u_1)T(u_2)\bar{T}(\bar{u}_1)}\\
=&\int d^2u_1[2\pi i\p_{\tau}\vev{T\bar{T}(u_1)T(u_2) }+\vev{T}\vev{T\bar{T}(u_1)T(u_2) }+\frac{c}{12}(P''(w-u_1)+P''(w-u_2)\vev{T\bar{T}}\\
&+2(P(w-u_1)+2\eta)\vev{T\bar{T}(u_1)T(u_2) }+2(P(w-u_2)+2\eta)\vev{T\bar{T}(u_1)T(u_2) }\\
&+ (\zeta(w-u_1)+2\eta u_1)\p_{u_1}\vev{T\bar{T}(u_1)T(u_2) } +(\zeta(w-u_2)+2\eta u_2)\p_{u_2}\vev{T\bar{T}(u_1)T(u_2) }]
\ea\ee
with the function which has already computed in eq.(\ref{tbtt})
\be\ba
&\vev{T\bar{T}(u_1)T(u_2) }\\
=&\frac{8i\pi^3\p^2_\tau\p_{\bar{\tau}}Z}{Z}+2(P(u_1-u_2)+2\eta)(4\pi^2)\frac{\p_\tau\p_{\bar{\tau}}Z}{Z}+\frac{c}{12}P''(u_1-u_2)(-2\pi i)\p_{\bar{\tau}}\ln Z.
\ea\ee
Now we would like to compute each term in the RHS of eq.(\ref{TTTTar}).
Note the last term of eq.(\ref{TTTTar}) vanishes  since $\int d^2u_1P'(u_1-u_2)=0=\int d^2u_1P'''(u_1-u_2)$.

The first term of eq.(\ref{TTTTar}) is
\be\ba
2\pi i\int d^2u_1 \p_{\tau}\vev{T\bar{T}(u_1)T(u_2) }=&2\pi i\p_{\tau}\int d^2u_1 \vev{T\bar{T}(u_1)T(u_2) }\\
=&2\pi i\p_\tau\Big(-\frac{(2\pi i)^3\tau_2\p^2_\tau\p_{\bar{\tau}}Z}{Z}+(2\pi )^3\frac{\p_\tau\p_{\bar{\tau}}Z}{Z}\Big).
\ea\ee
The second term of eq.(\ref{TTTTar}) is
\be\ba
\int d^2u_1\vev{T}\vev{T\bar{T}(u_1)T(u_2) }=\Big(-\frac{(2\pi i)^3\tau_2\p^2_\tau\p_{\bar{\tau}}Z}{Z}+(2\pi )^3\frac{\p_\tau\p_{\bar{\tau}}Z}{Z}\Big)\vev{T}.
\ea\ee
The third term of eq.(\ref{TTTTar}) is
\be\ba
\int d^2u_1\frac{c}{12}(P''(w-u_1)+P''(w-u_2))\vev{T\bar{T}}=\frac{c}{12}P''(w-u_2)\tau_2\vev{T\bar{T}}.
\ea\ee
The fourth term of eq.(\ref{TTTTar}) is
\be\ba
&\int d^2u_1  2(P(w-u_1)+2\eta)\vev{T\bar{T}(u_1)T(u_2) }\\
=&\int d^2u_1  2(P(w-u_1)+2\eta)\\
&\times[\frac{8i\pi^3\p^2_\tau\p_{\bar{\tau}}Z}{Z}+2(P(u_1-u_2)+2\eta)(4\pi^2)\frac{\p_\tau\p_{\bar{\tau}}Z}{Z}+\frac{c}{12}P''(u_1-u_2)(-2\pi i)\p_{\bar{\tau}}\ln Z]\\
=&2\pi\frac{8i\pi^3\p^2_\tau\p_{\bar{\tau}}Z}{Z} +(4\pi^2)\frac{\p_\tau\p_{\bar{\tau}}Z}{Z}(4P_{w,u_2}+16\eta^2\tau_2+8\eta(\pi-2\eta\tau_2))
 +\frac{c}{6}(-2\pi i)P''_{w,u_2}\p_{\bar{\tau}}\ln Z \\
&= \frac{16i\pi^4\p^2_\tau\p_{\bar{\tau}}Z}{Z} +(4\pi^2)\frac{\p_\tau\p_{\bar{\tau}}Z}{Z}(4P_{w,u_2} +8\eta \pi )
 +\frac{c}{6}P''_{w,u_2}(-2\pi i)\p_{\bar{\tau}}\ln Z,
\ea\ee
where we introduce the notation $P_{w,u_2}=\int d^2u_1P(u_1-w)P(u_1-u_2)$  and $P''_{w,u_2}=\int d^2u_1P(u_1-w)P''(u_1-u_2)$ which are computed below in (\ref{pab}) and (\ref{pppab}) respectively.

The fifth term of eq.(\ref{TTTTar}) is
\be\ba
&\int d^2u_12(P(w-u_2)+2\eta)\vev{T\bar{T}(u_1)T(u_2) }
\\
=&2(P(w-u_2)+2\eta)\Big(-\frac{(2\pi i)^3\tau_2\p^2_\tau\p_{\bar{\tau}}Z}{Z}+(2\pi )^3\frac{\p_\tau\p_{\bar{\tau}}Z}{Z}\Big).
\ea\ee
The sixth term of eq.(\ref{TTTTar}) is
\be\ba \label{sixth}
 &\int d^2u_1(\zeta(w-u_1)+2\eta u_1)\p_{u_1}\vev{T\bar{T}(u_1)T(u_2) }\\
=&\int d^2u_1(\zeta(w-u_1)+2\eta u_1)\p_{u_1}\Big[2P(u_1-u_2)(4\pi^2)\frac{\p_\tau\p_{\bar{\tau}}Z}{Z}+\frac{c}{12}P''(u_1-u_2)(-2\pi i)\p_{\bar{\tau}}\ln Z\Big].
\ea\ee
which can be computed as follows. Firstly, consider the following integral
\be \ba
&\int d^2u_1\zeta(w-u_1)\p_{u_1}P(u_1-u_2)\\
=&\int d^2u_1[\p_{u_1}(\zeta(w-u_1)P(u_1-u_2))-P(u_1-u_2)\p_{u_1}\zeta(w-u_1)]\\
=&\int d^2u_1[\p_{u_1}(\zeta(w-u_1)P(u_1-u_2))] -P_{w,u_2},
\ea\ee
where the second term is defined and computed in eq.(\ref{pab}) as mentioned before, while the first term is
\footnote{
In the last step, the following integral is along the real axis since $2w_1=1$, so $d\bar{u}_1=du_1$
\be
\int_0^1du_1 P(u_1-u_2)=-\int_0^1du_1 \p_{u_1}\zeta(u_1-u_2)=-2\eta.
\ee
  To  evaluate the second term in the last line of eq.(\ref{int2}), we parametrized the integral path as (notice $2w_2=\tau=\tau_1+i\tau_2$)
\be
du_1=\Big(1+i\frac{\tau_2}{\tau_1}\Big)dt,~~d\bar{u}_1=\Big(1-i\frac{\tau_2}{\tau_1}\Big)dt,~~~~t\in(0,\tau_1],
\ee
and
\be
P(u_1-u_2)=P\Big(\Big(1+i\frac{\tau_2}{\tau_1}\Big)t-u_2\Big)=-\p_{u_1}\zeta(u_1-u_2)=-\frac{dt}{du_1}\p_t \zeta\Big(\Big(1+i\frac{\tau_2}{\tau_1}\Big)t-u_2\Big).
\ee
Then
\be\ba
\int_0^{2w_2} d\bar{u}_1P(u_1-u_2)=&- \frac{dt}{du_1}\Big(1-i\frac{\tau_2}{\tau_1}\Big)\int_0^{\tau_1}dt\p_t \zeta\Big(\Big(1+i\frac{\tau_2}{\tau_1}\Big)t-u_2\Big)\\
=&- \frac{dt}{du_1}\Big(1-i\frac{\tau_2}{\tau_1}\Big) \zeta\Big(\Big(1+i\frac{\tau_2}{\tau_1}\Big)t-u_2\Big)\Big|_0^{\tau_1}
= \frac{-2\eta'\bar{\tau}}{\tau}.
\ea\ee}
\be\ba \label{int2}
&\int d^2u_1\p_{u_1}(\zeta(w-u_1)P(u_1-u_2))\\
=&-\frac{i}{2}\oint_{\p T^2} d\bar{u}_1 \zeta(u_1-w)P(u_1-u_2)\\=&-\frac{i}{2}(-2\eta')\int_0^{2w_1}d\bar{u}_1P(u_1-u_2)-\frac{i}{2}2\eta\int_0^{2w_2} d\bar{u}_1P(u_1-u_2)\\
=&-2i\eta\eta'+2i\frac{\bar{\tau}}{\tau}\eta\eta'.
\ea\ee
Similarly we can compute the remaining integrals in eq.(\ref{sixth}), which are
\be \ba
&\int d^2u_1u_1\p_{u_1}P(u_1-u_2)\\
=&\int d^2u_1[\p_{u_1}(u_1P(u_1-u_2))-P(u_1-u_2)]\\
=&  i\tau\eta - i \eta'\frac{\bar{\tau}}{\tau}-(\pi-2\eta\tau_2 )
= i\bar{\tau}\eta - i \eta'\frac{\bar{\tau}}{\tau}-\pi
\ea\ee
and
\be \ba
&\int d^2u_1(\zeta(w-u_1)+2\eta u_1)\p_{u_1}
P''(u_1-u_2) =-\int d^2u_1 P(w-u_1)P''(u_1-u_2)=-P''_{w,u_2}.
\ea\ee
Therefore the  eq.(\ref{sixth}) is
\be\ba
 &\int d^2u_1(\zeta(w-u_1)+2\eta u_1)\p_{u_1}\vev{T\bar{T}(u_1)T(u_2) }\\
=&\int d^2u_1(\zeta(w-u_1)+2\eta u_1)\p_{u_1}\Big[2P(u_1-u_2)(4\pi^2)\frac{\p_\tau\p_{\bar{\tau}}Z}{Z}+\frac{c}{12}P''(u_1-u_2)(-2\pi i)\p_{\bar{\tau}}\ln Z\Big]\\
=&2(4\pi^2)\frac{\p_\tau\p_{\bar{\tau}}Z}{Z}\Big(-2i\eta\eta'+2i\frac{\bar{\tau}}{\tau}\eta\eta'-P_{w,u_2}+2\eta( i\bar{\tau}\eta - i \eta'\frac{\bar{\tau}}{\tau}-\pi) \Big)+\frac{c}{12}(2\pi i)P''_{w,u_2}\p_{\bar{\tau}}\ln Z\\
=&2(4\pi^2)\frac{\p_\tau\p_{\bar{\tau}}Z}{Z}\Big(-2i\eta\eta' -P_{w,u_2}+2i \bar{\tau}\eta^2   -2\pi\eta \Big)+\frac{c}{12}(2\pi i)P''_{w,u_2}\p_{\bar{\tau}}\ln Z.
\ea\ee
Finally, collecting all terms together, eq.(\ref{TTTTar}) equals
\be \ba
&\int  d^2u_1\vev{T(w)T(u_1)T(u_2)\bar{T}(\bar{u}_1)}\\
=&2\pi i \Big( \frac{3(2\pi )^3 \p^2_\tau\p_{\bar{\tau}}Z}{2Z}-\frac{(2\pi i)^3\tau_2\p^3_\tau\p_{\bar{\tau}}Z}{Z}\Big)+\frac{c\tau_2}{12}P''(w-u_2)\vev{T\bar{T}} \\
&+ \frac{16i\pi^4\p^2_\tau\p_{\bar{\tau}}Z}{Z} +\frac{(16\pi^2)\p_\tau\p_{\bar{\tau}}Z}{Z}( P_{w,u_2} +2\eta \pi )
 \\&+2(P(w-u_2)+2\eta )\Big(-\frac{(2\pi i)^3\tau_2\p^2_\tau\p_{\bar{\tau}}Z}{Z}+(2\pi )^3\frac{\p_\tau\p_{\bar{\tau}}Z}{Z}\Big) \\
 &+ (8\pi^2)\frac{\p_\tau\p_{\bar{\tau}}Z}{Z}\Big(-2i\eta\eta' -P_{w,u_2}+2i \bar{\tau}\eta^2   -2\pi\eta \Big).
\ea\ee
\subsection{Computation of $P_{a,b},P''_{a,b}$}
In the following we will calculate the integrals
\be
P_{a,b}\equiv\int d^2zP(z-a)P(z-b),~~P''_{a,b}\equiv\int d^2zP(z-a)P''(z-b).
\ee
Firstly, consider $P_{w,u_2}$,  of which the integrand is elliptic, thus it can be expressed in terms of $\zeta$ function and its derivatives, according to the position and order of the poles \cite{ellipticbook}. More precisely, since $P(z-a)P(z-b)$ has order two poles at $z=a$ and $z=b$ respectively, then we can write
\be \ba \label{PP}
P(z-a)P(z-b)=&a_0+a_1\zeta(z-a)+b_1\zeta(z-b)-a_2\zeta'(z-a)-b_2\zeta'(z-b).
\ea\ee
with $a_i,b_i$ are constants which can be determined by comparing the coefficients of the poles in both side and so on. It turns out that those constants are
\be
a_2=b_2=P(a-b)
,~~
a_1=-b_1=-P'(a-b),
\ee
and
\be
a_0=P(a)P(b)-P'(a-b)(\zeta(a)-\zeta(b))-P(a-b)(P(a)+P(b)).
\ee
To evaluating the integral of the RHS of eq.(\ref{PP}), we integrate each term separately. For the second and third terms which take the form $\zeta(z-a)-\zeta(z-b)$ we have to consider the following integral
\be \ba
B(a)\equiv &\int  d^2z\zeta(z-a)=\int d^2z(\ln\sigma(z-a))'=\frac{i}{2}\oint d\bar{z}\ln \sigma (z-a)\\
=&\frac{i}{2}\int_0^{2w_1}d\bar{z}(\ln \sigma (z-a)-\ln \sigma (z-a+2w_2))\\
&+\frac{i}{2}\int_0^{2w_2}d\bar{z}(\ln \sigma (z-a+2w_1)-\ln \sigma (z-a))\\
=&\frac{i}{2}\int_0^{2w_1}d\bar{z}(-1)(\pi i+2\eta'(z-a+w_2))+\frac{i}{2}\int_0^{2w_2}d\bar{z}(\pi i+2\eta(z-a-w_1)),
\ea\ee
then
\be
B(a)-B(b)=i(a-b)(\eta'-\eta\bar{\tau})=(a-b)(\pi-2\eta \tau_2).
\ee
Finally we obtain
\be\ba \label{pab}
P_{a,b}=\int d^2 zP(z-a)P(z-b)=a_0\tau_2+a_1(a-b)(\pi-2\eta \tau_2)+2a_2(\pi-2\eta \tau_2).
\ea\ee
Next consider $P''_{a,b}$ whose integrand $P(z-a)P''(z-b)$ is also elliptic. Following the same steps as above, firstly we express the integrand in terms of $\zeta$ function and its derivatives, which can be achieved by taking derivative on eq.(\ref{PP})  with respect to $b$ twice. Then we integrate the resulting expression, which turns out to be
\be\ba \label{pppab}
P''_{a,b}=&\int d^2 zP(z-a)P''(z-b)=a''_0\tau_2+a''_1(a-b)(\pi-2\eta \tau_2) - a'_1(\pi-2\eta \tau_2),
\ea\ee
where the prime on $a_i$ denotes the derivatives with respect to $b$.
Note the RHS of eq.(\ref{pppab}) can also be obtained by directly taking derivative on RHS of eq.(\ref{pab})  twice with respect to $b$ twice.

\end{document}